%% file: main.tex
\documentclass[sigconf]{acmart}
\usepackage{subfigure}

\setcopyright{none}

\begin{document}
\title{Wireless Charging Power Side-Channel Attacks} 
\author{Alexander S. La Cour, Khurram K. Afridi, and G. Edward Suh\\
Cornell University, Ithaca, NY\\
\{asl247, kka34, gs272\}@cornell.edu}

\begin{abstract}
This paper shows that today's wireless charging interface is vulnerable to power side-channel attacks;
a smartphone charging wirelessly leaks private information about its activity to the wireless charger (charging transmitter). 
We present a website fingerprinting attack through the wireless charging side-channel for both iOS and Android devices.
The attack monitors the current drawn by the wireless charging transmitter while 20 webpages from the Alexa top sites list are loaded on a charging smartphone. We implement a classifier that correctly identifies unlabeled current traces with an accuracy of 87\% on average for an iPhone 11 and 95\% on average for a Google Pixel 4.
This represents a considerable security threat because wireless charging does not require any user permission if the phone is within the range of a charging transmitter. 
To the best of our knowledge, this work represents the first to introduce and demonstrate a power side-channel attack through wireless charging.
Additionally, this study compares the wireless charging side-channel with the wired USB charging power side-channel, showing that they are comparable.
We find that the performance of the attack deteriorates as the contents of websites change over time.
Furthermore, we discover that the amount of information leakage through both wireless and wired charging interfaces heavily depends on the battery level; minimal information is leaked at low battery levels. 
\end{abstract}


\maketitle

\input{introdution}
\input{background}
\input{overview}

\input{fingerprinting_attack}
\input{results}
\input{countermeasure}
\input{related}
\input{conclusion}
\begin{acks}
This work was supported in part by the Semiconductor Research Corporation.
\end{acks}

\bibliographystyle{ACM-Reference-Format}
\bibliography{main}

\end{document}

%% file: introdution.tex
\section{Introduction}
\label{sec:intro}

Smartphone usage and charging have become increasingly prevalent. According to a Pew Research Center survey, 81\% of American adults report owning a smartphone~\cite{pew}. Moreover, a market research poll conducted by Veloxity, a phone charging station company, found that on average, respondents charged their phones from 1.6 to 2.7 times per day~\cite{vel}. While wired chargers are currently more common, the market share of wireless charging solutions has been consistently expanding and wireless chargers are supplanting wired chargers. A BIS research report claims the global wireless charging market will be worth over \$20.97B in 2023, and the CEO of BIS Research has claimed that will be more wireless chargers than cables by that point~\cite{bis}.

In this paper, we show that today's wireless charging interfaces are vulnerable to a power side-channel attack 
that can leak private information from a charging device to a wireless charger (charging transmitter). 
In particular, we demonstrate the attack on the Qi standard~\cite{qi}, which is currently the dominant standard for wireless charging.
The side-channel attack through wireless charging represents an important threat because it does not require a physical connection to a victim device, and can occur without user permission or any sophisticated equipment.
While a similar power side-channel attack has been previously demonstrated through the traditional wired charging interface, wireless charging has been considered noisy and therefore secure against practical side-channel attacks.
This paper is the first to investigate power side-channel attacks through wireless charging and demonstrate that practical attacks are feasible.

As a concrete example, we study a website fingerprinting attack through the wireless charging power side-channel, and perform detailed experimental studies on an Apple iPhone 11 and a Google Pixel 4. 
The phones are placed on a wireless charging transmitter and loads a webpage from a set of 20 candidates. While the webpage is loaded on the phone, the amount of current being drawn by the wireless charging transmitter is recorded as a current trace. 
We find that after collecting enough current traces, it is possible to train a classifier to correctly identify the webpage that was loaded at the time an unlabeled current trace was recorded. We were able to achieve an accuracy of at least 80\% on average with current traces as short as 2.5 seconds on the Qi Baseline Power Profile.

Our study also shows that this power side-channel attack can be performed without expensive or bulky measurement equipment such as a high-performance oscilloscope, which
makes concealing a power monitoring circuit at a malicious wireless charger quite plausible. 
In our experimental setup, we used a small microcontroller to measure the current at a charger. 
We believe that the attack circuits can be even more tightly integrated with an existing microcontroller on a malicious charger. 
When a smartphone owner uses a public wireless charging station, they will generally not have access to the circuitry of the wireless charging transmitter and will not be able to identify a malicious charger.
Furthermore, public charging stations are becoming ubiquitous and are increasingly supporting Qi-enabled wireless charging. There are currently over 190 smart devices that natively support the Qi standard, and older phones can implement the standard by connecting to a Qi compatible wireless receiver via an accessory or case for as little as \$10.
Given the prevalence of wireless charging and the ease of an attack, we believe that the side-channel attack through wireless charging represents a significant security risk.

\begin{figure*}[!t]
    \centering
    \includegraphics{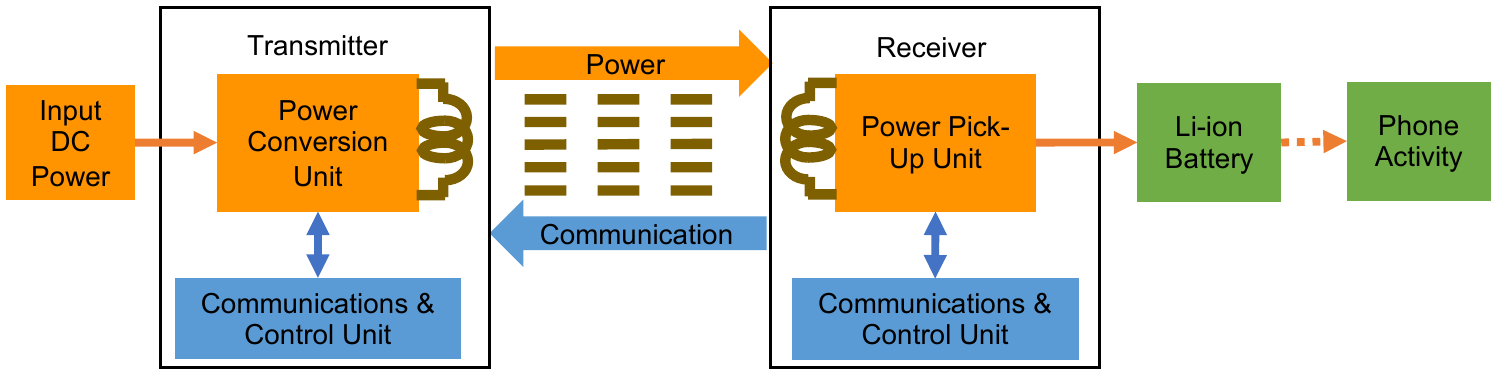}
    \caption{A diagram of the transmitter and receiver hardware for the Qi standard.}
    \label{fig:qi-hardware}
\end{figure*}

In addition to demonstrating that today's wireless charging interface is vulnerable to practical power side-channel attacks, this paper also presents the results from a set of in-depth experimental studies in order to better understand the capabilities and limitations of the wireless charging side-channel.
For example, we compare wireless charging and traditional wired charging in the context of power side-channel attacks. 
The experimental results show that the wireless charging side-channel, while noisy, is comparable to the wired side-channel, leading to similar or better website prediction accuracy depending upon the device attacked. We also observed the effects of other variables including the length of time between the collection of training and testing traces, and the amount of power delivered by the charger.

Our study also found that the information leakage through these side-channels in today's battery-powered devices depends heavily on the state of charge of the battery. 
When the battery level is high, the power consumption of the victim device is almost directly reflected on the power draw from the charger, revealing the activities on the device.
On the other hand, when the battery level is low, most of the power from a charger is used to charge the battery.
In that sense, we found that devices are far more vulnerable to wireless charging side-channel attacks when the battery level is above 80\%. 
Unfortunately, given their convenience, users often leave devices on wireless chargers when fully charged. 
The chairman of the Wireless Power Consortium (WPC) stated that the WPC was unaware of negative consequences of prolonged wireless charging and suggested that topping off a phone battery will increase its life span~\cite{nyt}.
For user privacy, our study suggests that future devices may want to adjust their charging algorithm and avoid fully charging a battery through an untrusted wireless charger.

The following summarizes the main technical contributions of this paper.
\begin{itemize}

\item This paper represents the first demonstration of the existence of a wireless charging power side-channel on today's smartphones. Even with noise, the wireless charging power side-channel leaks enough information to allow accurate website fingerprinting on a charging smartphone. 

\item This paper experimentally compares the wireless and wired charging side-channels, and shows that they leak the same power consumption information. Additionally, traces from the wireless and wired charging can be used to classify each other.

\item This paper shows that the amount of information leaked through these side-channels depends significantly on battery level. The exact amount of information leakage at different battery levels depends on the model of the charging device.

\end{itemize}

The rest of this paper is organized as follows: Section 2 discusses background information related to wireless charging, the Qi standard, and the concept of power side-channel attacks. Section 3 introduces our threat model and presents high-level observations about the wireless charging power side-channel along with our experimental setup. Section 4 provides an overview of our website fingerprinting attack and presents our classification algorithm, Section 5 details the experimental results of the attack and the impact of a number of variables. Section 6 discusses possible countermeasures, limitations, and future research directions. Related work is discussed in Section 7 and Section 8 concludes the paper.

%% file: background.tex
\section{Background}
\label{sec:background}

This section provides general technical background on wireless 
charging and power side-channel attacks which is necessary to understand
the proposed wireless charging power side-channel attack.

\subsection{Wireless Charging}

Using the open interface standard Qi for wireless power transfer is the prevailing method for wirelessly charging smart devices. Qi was developed by the Wireless Power Consortium and describes the functional and physical characteristics necessary to allow the exchange of power and information between a receiver and a transmitter. Currently, Qi supports two power specifications to charge mobile devices: the Qi Baseline Power Profile, which delivers power below 5 W, and the Qi Extended Power Profile, which supports up to 15 W~\cite{wpcs}. Wireless charging has quickly become standard in new devices; following its release in 2008, Qi was integrated into over 200 smart devices by 2021~\cite{qi}. The ubiquity of wireless charging in the form of public charging stations makes the consequences of potential side-channel attacks severe.

Qi utilizes inductive charging to wirelessly transfer power from a transmitter to a receiver. Under this charging scheme, an induction coil on the transmitter (the primary coil) couples to another coil on the receiver (the secondary coil). The transmitter then runs an alternating current through its coil which induces a charge in the receiving coil by Faraday's law of induction. Additionally, resonant inductive coupling is employed so that the devices can charge while up to 4 cm apart. Resonant inductive coupling occurs in coupling systems that have capacitors connected to both induction coils, creating LC circuits with individual resonance frequencies~\cite{wpcs}. The alternating current driven by the transmitter can then cause the load-bearing side to resonate which increases the coupling strength. The current in the receiving coil is then rectified into direct current so that it can be employed to charge a battery or directly power a device.

The implementation of the Qi standard requires special circuitry that interposes between the power source and the device battery. Figure~\ref{fig:qi-hardware} shows that communication between the two devices occurs via backscatter modulation and is unidirectional from the receiver to the transmitter. The transmitting coil is a part of a power conversion unit while the receiving coil is a part of a power pick-up unit. Both the transmitter and receiver contain communications and control units that use PID controllers in order to balance the transferred power level to the amount requested by the charging device. 

The communication protocol of the Qi standard involves five phases. In the first phase, the power transmitter sends out an analog ping to detect whether or not an object is present. The power transmitter then sends out a longer, digital ping in order to give the receiver time to reply with a signal-strength packet. If the transmitter determines this packet is valid, it will continue to power its coil and proceed to the next step. The third phase is known as the identification and configuration phase, where information is sent by the receiver in packets in order to properly configure the transmitter for power transfer. Next, the power-transfer phase begins, during which the receiver sends control error packets to modify the power supply. The final phase occurs when the power receiver stops communication or specifically requests the end of power transfer~\cite{wpcs}.

In terms of power delivery, Qi wireless charging is less efficient than wired charging. Wireless charging introduces noise, and some have speculated that this type of noise is a good countermeasure against side-channel attacks that examine the amount of current used to charge a smartphone~\cite{daily_swig}. However, wireless charging transmitters do not store any significant amount of charge. Therefore, most current that enters the transmitter will directly reflect the phone activity which acts as a load on the receiver.

\vspace{-0.1in}
\subsection{Battery Charging Cycles}
Most smartphones use lithium-ion (Li-ion) batteries. These batteries go through different charging stages~\cite{microchip}. The first stage, known as constant current, involves supplying the maximum current to the battery, steadily increasing its voltage. Once the voltage of the battery reaches approximately 4.2 V, the second stage, known as the constant voltage stage, will begin. During this phase, the supplied current drops off in order to maintain the current voltage level of the battery. After the battery state of charge has reached 100\%, if it is still charging, the charger will provide a topping charge to make up for any phenomena that discharged the battery and return the state of charge to 100\%~\cite{time}.

As a result of the charging stages, the amount of current drawn by a phone heavily depends on the battery state of charge regardless of how much power the device is consuming. When the phone battery is at a low state of charge that corresponds to the constant current stage, the amount of power a phone is consuming will not significantly affect its overall current draw. This is because the phone is consuming power from the battery, and as long as the battery remains below the threshold for constant voltage to be applied, the same amount of maximum current will be delivered in order to charge the battery. On the other hand, when the battery of the smartphone is in the constant voltage stage, the power consumption of the phone will affect the voltage of the battery and the current will vary in order to maintain the desired voltage. When the phone battery is fully charged, the amount of power drawn from the battery is a direct reflection of the amount of current supplied to top off its charge.

\vspace{-0.1in}
\subsection{Power Side-Channels}

Side-channel attacks are methods to acquire sensitive information through unintended secret-dependent variations in physical behaviors. 
The information leaked from a side-channel attack is a byproduct of computations occurring on hardware and is not a specific software vulnerability. 
Power side-channel attacks are a specific type of side-channel attack that analyze the power traces of the electrical activity on a device to extract information~\cite{kocher}. Simple power analysis is a method of power side-channel attack that infers a secret value from a power trace by identifying power consumption profiles that directly depend on the secret. Frequency filters and averaging functions can be applied to filter out noise in these power traces~\cite{clark-identify}. Differential power analysis is a more complex method of side-channel attack that allows identification of intermediate values within cryptographic computations after a statistical analysis of data collected prior. Signal processing and error-correcting can also be applied to DPA attacks. 

While power side-channel attacks are an established field of research, applying these techniques to mobile devices is a relatively new endeavor. Mobile devices are uniquely susceptible to side-channel attacks because they are portable, generally powered on, and have a multitude of sensors. Understanding the extent of sensitive information that a power side-channel attack can infer will provide insight into security risks.

Smartphone security relies on two basic premises, application sandboxing and a permission system. These rules ensure that applications cannot access sensitive information contained in another resource. Yet, even without direct access to the data pins of the smart device, power side-channel attacks have proven to be effective against smartphones. A public USB power station for a smartphone can be considered as a potential physical adversary for a power side-channel because it requires the phone to have a direct connection to the station that is collecting data on its power usage. This type of attack is non-invasive because it does not manipulate the packaging of the chip and is passive because the power consumption is only observed and not influenced. 

\begin{figure}[!h]
  \centering
  \includegraphics{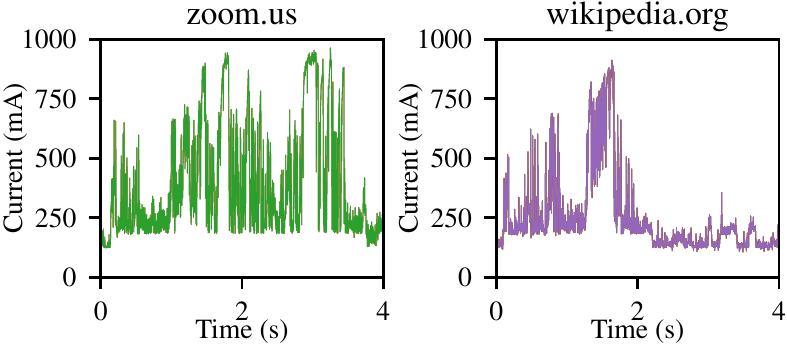}
  \caption{A wireless charger draws a varying amount of current as mobile webpages are loaded on the charging phone.}
  \label{wired}
\end{figure}

For example, Yang et al.~\cite{yang-usb} showed that charging a smartphone over a USB cable exposes a side-channel that is vulnerable to an SPA attack. By monitoring the power that a charging smartphone drew while loading webpages, they were able to successfully infer private browsing information. Figure~\ref{wired} shows that in the current traces we collected, different websites leave unique signatures through the wireless charging side-channel over short time durations.

%% file: overview.tex
\section{Power Side-Channels in Wireless Charging}
\label{sec:overview}

This section introduces the concept of wireless charging power side-channel attacks and discusses their capabilities and limitations at a high-level. 
The following section provides more in-depth study using website fingerprinting as a concrete example attack.

\subsection{Threat Model}
\begin{figure}[!h]
  \centering
  \includegraphics{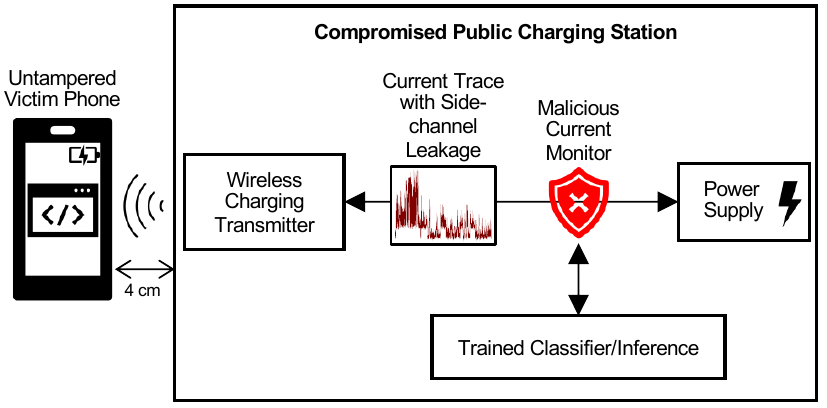}
  \caption{Threat model demonstrating a power side-channel attack by a compromised public charging station.}
  \label{fig:threatmodel}
\end{figure}

Figure~\ref{fig:threatmodel} shows the threat model that is assumed for the wireless charging side-channel attack. Under this threat model, an attacker is able to monitor and record the amount of power being delivered to an untampered Qi wireless transmitter from a malicious public wireless charging station. The target device performs activities that depend on sensitive events or data values, which influence its power consumption. The goal of the attacker is to infer these events or data values on the target device by analyzing the recorded power traces. While we assume the public charging station is compromised, it need not be malicious because the classification and inference can occur remotely.

Wireless charging does not require any user permissions or initiation and will begin immediately if both the mobile device and the transmitter follow the Qi standard and are in range (4 cm). There is no need for the device to be plugged in to the charging station. The target device is not assumed to have any malicious software and this threat model does not depend on any particular software vulnerability. Additionally, this type of attack does not require any physical tampering of the target device or battery.

\subsection{Experimental Setup}

The high-level idea of the wireless power side-channel attack is similar to that of the traditional wired charger power side-channel attack.
However, given that wireless charging interfaces do not have physical wire connections and are likely to be more susceptible to noise, it has been hypothesized~\cite{daily_swig} that power side-channel attacks will not be practical through wireless charging. 
In that sense, the main technical contributions of this paper lie in experimental studies that demonstrate that wireless power side-channel attacks are feasible in today's mobile phones and their capabilities are comparable to those of wired power side-channel attacks.

Here we briefly describe the experimental setup that we used. The experiments are designed to understand the capabilities and limitations of the wireless power side-channels:

\begin{itemize}
    \item Does the wireless power side-channel leak enough information to infer activities on a mobile device even with noise in the wireless interface? Are the measurements repeatable? 
    \item How is the wireless power side-channel impacted by the battery level? 
    \item How does the wireless power side-channel compare to the wired power side-channel in terms of leakage?
\end{itemize}

\begin{figure}[!h]
    \centering
    \subfigure[Overview of current trace monitoring.]{\includegraphics[scale=.95]{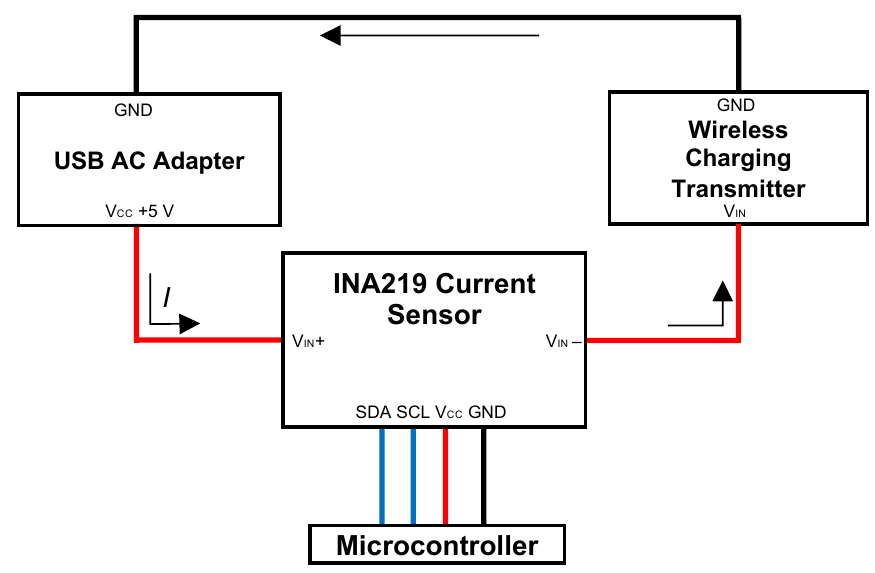}}
    \hfill
    \subfigure[Photo of setup with the Adafruit 5 W transmitter.]{\includegraphics[scale=.95]{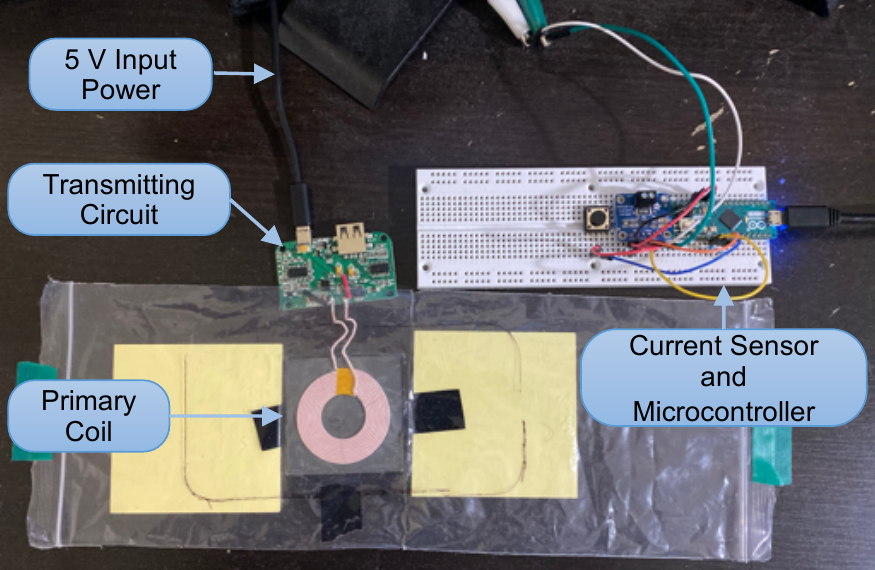}}
    \caption{Current trace collection.}
    \label{fig:setup}
\end{figure}

\textbf{Current Trace Collection Circuit.} The DC current delivered to either a 5 W Adafruit Qi Wireless Charging Transmitter or a 10 W Max Anker Wireless Charging Pad from a USB AC adapter was sampled by placing an INA219 High Side DC Current Sensor in series with the V\textsubscript{CC} wire of the Micro-USB cable that charged the transmitters. This is depicted in Figure~\ref{fig:setup}. 

In order to collect wired current traces, the current sensor was instead placed in series with a spliced USB-A to Lightning or USB-A to USB-C cable. An Arduino Micro was then programmed to sample the current sensor at a frequency of 700 Hz (500 Hz in Sections 5.4 and 5.6). The cost of the entire current trace collection circuit used in this work is less than \$30.

\begin{figure}[!h]
    \centering
    \includegraphics{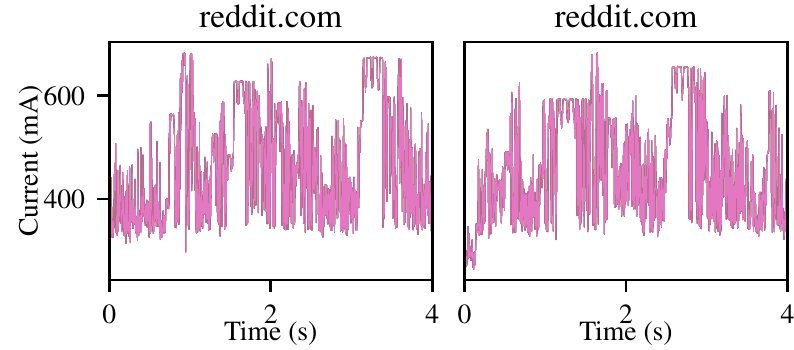}
    \includegraphics{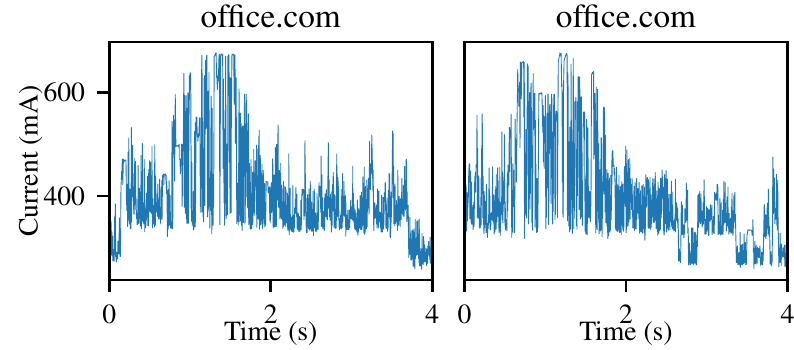}
    \includegraphics{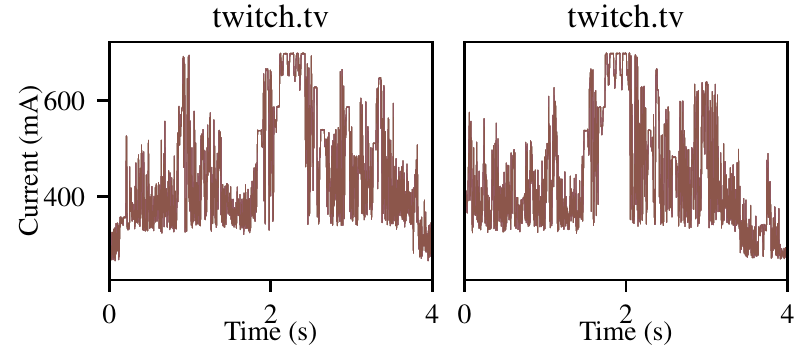}
    \caption{Current traces demonstrating the activity leaked when automatically loading webpages on an iPhone 11.}
    \label{fig:wireless_sites}
\end{figure}

\textbf{Example Current Traces.} Figure~\ref{fig:wireless_sites} demonstrates that like the USB charging side-channel, the wireless charging side-channel also leaks enough information to distinguish different websites. Additionally, we find that the collected current traces are repeatable across different trials indicating that the activity visible in the traces is a direct result of loading a particular website. In all cases, the websites take a variable amount of time to load and once fully loaded, the current drawn by the charging transmitter returns to a steady level.

\textbf{Phone Configuration.} The attack is performed on an Apple iPhone 11 (2019) running iOS 14  and a Google Pixel 4 (2019) running Android 11 which are both capable of wireless charging with Qi-certified chargers up to powers of 7.5 W and 11 W respectively. 

When the iPhone 11 traces were collected without noise, an outline for the phone was placed around the coil so that it could be positioned consistently above the transmitter across every trace. Otherwise, both phones were placed at various orientations while remaining centered enough to properly charge.

The phones used Wi-Fi to load the webpages. Several measures were taken to reduce the impact of the cache on reloading previously visited webpages including closing all other tabs and enabling private browsing. On the Safari browser with the iPhone 11, the options to reload the page from the origin, empty the cache, and ignore the resource cache were also enabled.

\subsection{Impact of Battery Level}

\begin{figure}[!h]
    \centering
    \includegraphics{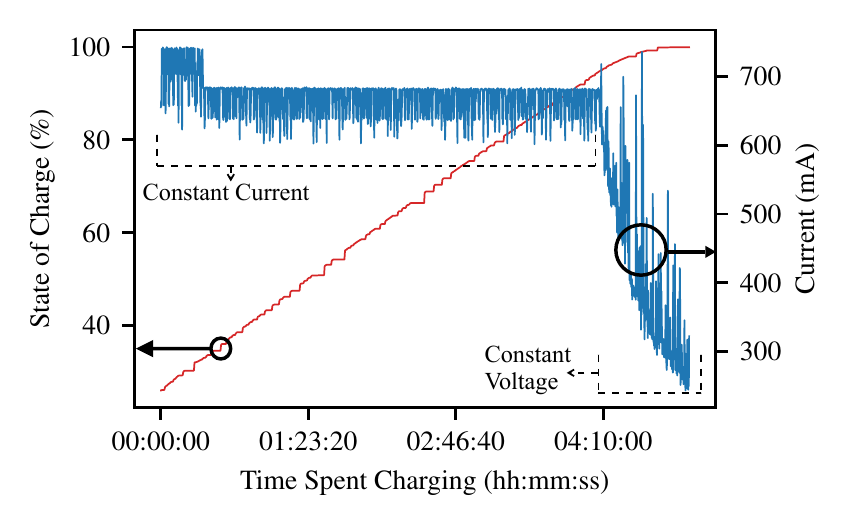}
    \caption{Current delivered by a 5 W Qi charger/battery state of charge vs charging time for an iPhone 11. The constant current and constant voltage charging stages are identified.}\label{fig:long}
    \centering
    \includegraphics{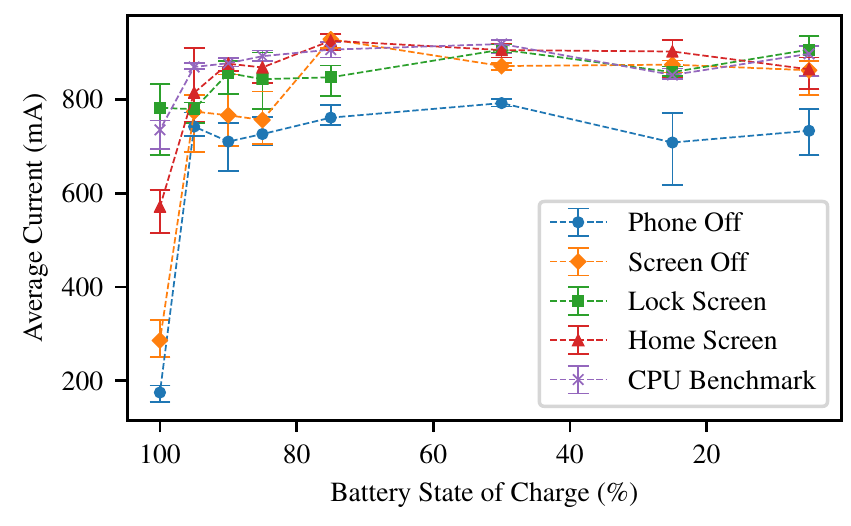}
    \caption{The average current consumption vs iPhone 11 state of charge for five different activities.}\label{fig:average-current}
\end{figure}

Figure~\ref{fig:long} shows how the wireless charger's current draw varies as the charging phone's battery level increases.
The red line represents the battery state of charge and the blue line shows the current draw.  
The results indicate that the charging profiles of a wireless charger mirror those of a wired charger~\cite{invio}.
At a low charge level, the current draw is relatively fixed except for a high-frequency component coming from the wireless interface. Then, the power draw gradually decreases as the battery state of charge increases.

Figure~\ref{fig:average-current} shows how the average current consumption of a wirelessly charging phone varies as it executes different processes. The experiment was carried out at 8 different battery levels.
While the results demonstrate that different processes do consume different amounts of power on average while wirelessly charging, a clear differentiation between activities only occurred while the phone's battery level is high. At battery states of charge less than or equal to 95\%, the activities were generally indistinguishable by the metric of average current assumption. The reason for this is that when the phone’s battery is fully charged, the amount of power delivered by the wireless transmitter is solely determined by the energy the phone is currently using as it cannot deliver more charge to a battery that is already at maximum capacity. If the battery is not fully charged, a part of the power delivered from the charger will also be used to charge the battery, regardless of the app running on the phone. 
    
Even if the average power consumption does not leak enough information to distinguish different activities at a lower battery level, a trace of dynamic power consumption over time can reveal far more information.
For example, in Section 5, we show that with the current traces over time, it is possible to distinguish different activities at battery levels lower than 95\%. For all experiments in our evaluation section, except for Section 5.6 where specific battery states of charge were examined, current traces were collected automatically beginning when the device's battery was full. During the duration traces were collected, the device's state of charge dropped to 90\%. In general, we found that battery-powered mobile devices are more susceptible to power side-channel attacks when the battery state of charge is high. The exact amount of information leaked depends on the charging algorithms used by a victim device. Our experiments in Section 5 suggest that even with time series data, the iPhone 11 leaks little information when the battery charge level is below 80\%.

%% file: fingerprinting_attack.tex
\section{Website Fingerprinting Attack}
In this section, our website fingerprinting attack is explained and the attack overview, data collection process, and classifier architecture are presented.

\subsection{Attack Overview}
The attacker seeks to utilize the collected power data to identify the webpages being loaded in a mobile browsing application by a victim as they charge their phone. As established by the mobile power side-channel attacks previously discussed, loading a website on a smartphone can affect its power consumption patterns. When the phone battery is near full charge, the power delivered to the wireless charging transmitter is directly proportional to the fluctuation in activity on the phone and will be recorded by the malicious public wireless charging station. 

A set of training data can be collected by the repeated loading of websites onto a charging device in this manner. This data can then be preprocessed and inputted to a website fingerprinting classifier for training and validation. After the model has been successfully trained, it can be used to classify new power data collected from victims by the malicious charging station. This victim data will then be similarly preprocessed to form the testing data, which if classified correctly, will reveal an individual's private browsing activity. This attack is performed on untampered wireless charging transmitters, but it is possible that a malicious transmitter that is designed for power side-channel attacks could provide more accurate traces.

\subsection{Current Trace Collection}

In the case of the iPhone 11, the Safari browser on the phone is connected to the Safari development tool, Web Inspector, on a Mac computer. The computer then runs a script that sequentially loads a set of websites on the iPhone 50 times. This is performed twice (once for wireless charging, once for wired charging) at each battery level examined. Trace collection on the Pixel 4 followed a similar process except that the Chrome browser and Chrome Developer Tools were used to initiate webpage loading. The current trace corresponding to the first 10 seconds of loading a website is recorded and between loading each site, the script waits 4 seconds. This script also automatically initializes the data collection in order to ensure that all power traces are synchronous and aligned. The top 20 websites from the Alexa Top Sites in United States list~\cite{alexa} were examined in this attack. 

For nearly all configurations, testing traces were collected with the intent to mirror normal device operation. This included setting the phone's brightness and volume at a constant level (although no websites visited automatically played audio), and enabling Bluetooth and cellular data. The exception to this is in Section 5.1, where test traces were collected with volume, Bluetooth, and cellular data disabled. For all traces, notifications on the devices were disabled in order to prevent calls from interrupting the data collection script. The Pixel 4 did not have a SIM card inserted, so it did not have cellular data enabled.

\subsection{Classification Algorithm}
\begin{figure*}[!t]
\centering
\includegraphics{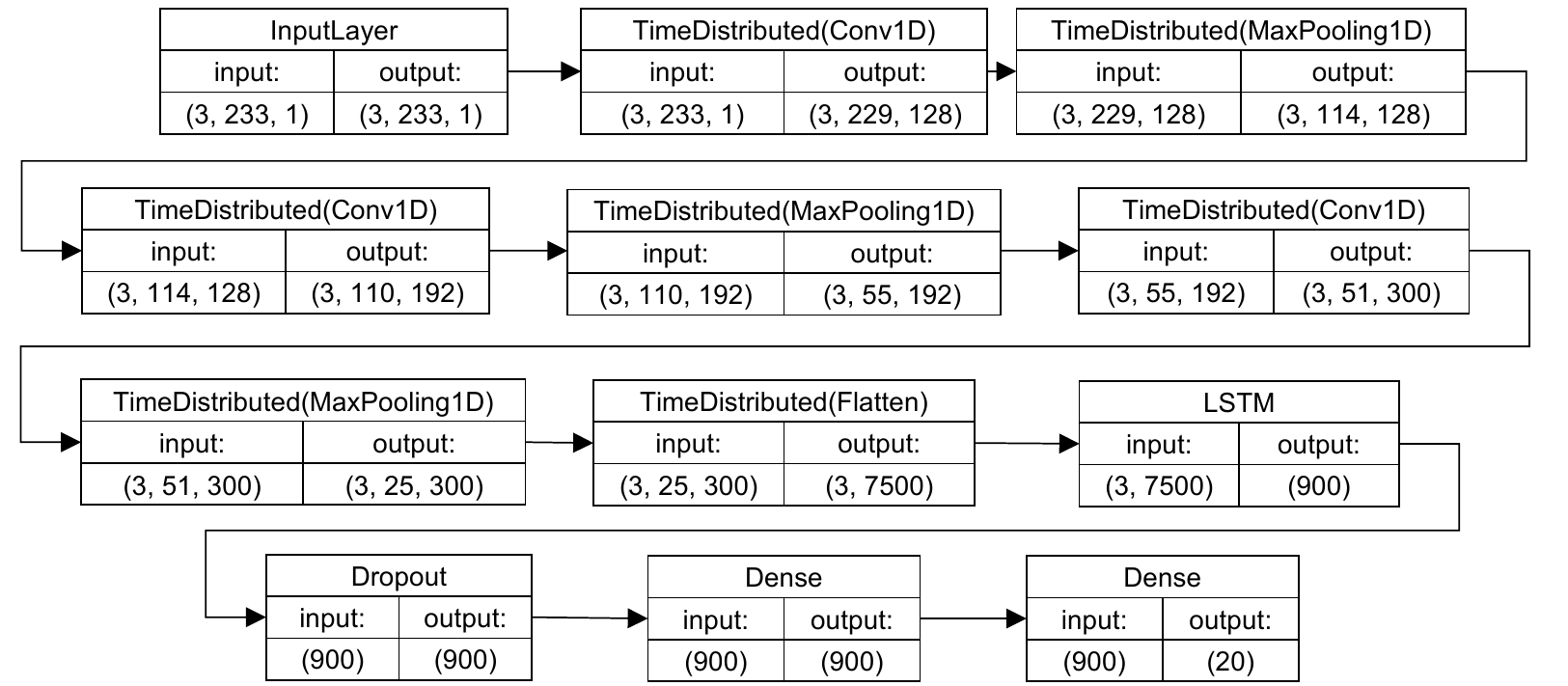}
\caption{1D CNN model where the duration of the windowed trace is 1 second. A layer labeled with (3, 233, 1) means that the layer can accept variable batch sizes of 1 second traces that have been segmented into three temporal slices of equal length to be processed separately by the convolutional layers and then interpreted together by the LSTM layer.}\label{fig:cnn-model}
\end{figure*}
For feature extraction, each current trace was broken into segments that represented 1 second of the original trace, with 97.5\% overlap. These segments were acquired by applying a sliding window algorithm to the overall current trace. This feature duration was chosen because many of the identifiable features that distinguished each trace were less than a second long. Training on many small segments rather than entire traces helped to increase the amount of training data available and reduced overfitting by making our model more shift-invariant. Each trace in the test set is broken into segments as was the training data, and each segment's classification is cast as a vote for classifying the overall test trace. The final trace label was assigned using a majority voting scheme. A 64/16/20 training/validation/testing split was used, which resulted in 200 unlabeled test current traces for each experiment conducted.

Deep neural networks act as both feature extractors and classifiers, which can make attacks more successful than traditional techniques. Additionally, convolutional neural networks (CNNs)~\cite{CNN,sadouk,cnn2} incorporate translation invariance, which allows them to recognize features even if they are translated to different time positions. Although our current traces were collected automatically, the loading time of pages sometimes is delayed randomly due to website traffic or other causes.

A 1D CNN, the architecture of which is pictured in Figure~\ref{fig:cnn-model}, is trained as a classifier on these segments and was implemented in the Keras~\cite{keras} software package. Our architecture is a modified version of a 1-D CNN that was used for human activity recognition~\cite{mlm}. This model was chosen as a base because it was designed for multi-output classification, had a foundational architecture that was easy to build on, and proved resilient to overfitting.

The topology of our CNN is three convolutional layers followed by a long short-term memory (LSTM) layer~\cite{lstm}, a fully connected layer, and a Softmax layer with 20 outputs, one for each website. Every convolutional layer used ReLU activation~\cite{relu}, had a convolutional window of size 5, and was followed by a max-pooling layer with a window of size 2 and a stride of 2. Each window was split into three equal length temporal slices in order to allow the LSTM layer to update its weights based on the chronological relationship it learned between the features from each slice. The CNN layers were wrapped in a TimeDistributed layer which is a layer that applies the same input operation across all time slices constructed from each window. There are 128 filters in the first convolutional layer, 192 in the second, and 300 in the third. The network also uses a dropout layer with a frequency of 50\% in order to further reduce overfitting by randomly dropping nodes and regularizing the network.

The LSTM layer was chosen for this classification problem because it is a recurrent neural network layer that is able to learn the order dependence within data. Given that the segments the network examines are 250-350 time steps in length, the ability of the classifier to learn order dependence would allow it to identify the presence of multiple features within a single segment. The data we collected was a one-dimensional time series and was a natural fit to this problem because while loading a website, many events such as executing JavaScript and loading images will always be executed by the phone in the same order. In this way, the LSTM layer complements the convolutional layers in our architecture: the convolutional layers extract features and the LSTM layer learns their order dependence.

The CNN outperformed all other classifiers we explored when evaluated on our collected data. The second best performance we obtained was with a Random Forest~\cite{randomforest} classifier that was trained with the frequency domain representation of the current traces. Although we were able to get reasonably high accuracy with this classifier, it did not perform as well on test traces that were translated when compared to training traces, and was not able to generalize to different charging conditions. In contrast, our CNN performed well on all scenarios in which current traces were collected and did not require any feature engineering aside from the application of the sliding window algorithm. Our model also successfully identified traces that were time-shifted with respect to the training data. In this way, our attack is proven to be able to conform to traces collected from multiple devices and charging methods with the same feature extraction method. This is critical because our threat model is intended to apply to a variety of phone models, operating systems, and chargers.

%% file: results.tex
\section{Evaluation}
In this section, we present the detailed experimental results on the website fingerprint attack
through wireless charging and discuss our findings. Rank 1 and Rank 2 identification accuracy of the classifier in different scenarios were calculated. Rank 1 counts a classification as correct if the majority vote picks the correct website for the trace. Rank 2 accuracy counts a classification as correct if either the website with the most or second-most votes is correct. The baseline accuracy of a random guess classifier for the 20 websites is 5\% for Rank 1 and 10\% for Rank 2.

We conducted a range of experiments aiming to identify how the classifier accuracy changed with respect to the following variables: (1) device manufacturer; (2) different devices for training and testing; (3) different chargers for training and testing; (4) noise; (5) length of current traces; (6) aging of training traces; (7) battery state of charge. The following subsections detail our findings and contributions with respect to each question.

\subsection{iPhone 11 vs Pixel 4}
In this subsection, we aim to identify how the accuracy of the classifier depends on the device used to collect current traces. The iPhone 11 and Google Pixel 4 were both used to collect current traces under the same conditions from both a 5 W wired charger and a 10 W wireless charger. Results from these experiments are reported in Table~\ref{table:r1} (iPhone) and Table~\ref{table:r2} (Pixel). All test traces in this section and in Section 5.2 included noise in the form of normal device operation conditions which included leaving the phones' Bluetooth, cellular data, volume, and notifications on while placing them at a variety of alignments with the transmitting coil.

The classifier achieved a Rank 1 accuracy of at least 82.0\% and a Rank 2 accuracy of at least 87.0\% when classifying wireless traces from the iPhone 11 with trace durations ranging from 2.5 to 6 seconds. Pixel 4 wireless traces were classified with more accuracy, especially at longer trace lengths. It achieved a Rank 1 accuracy of at least 85.5\% and a rank 2 accuracy of at least 91.5\% with trace durations ranging from 2.5 to 6 seconds. The high identification accuracy of the classifier in these scenarios indicates that the small changes in processor activity that occur when loading various websites are detectable through this wireless side-channel in both devices examined.

\begin{table}[!h]
\centering
\begin{tabular}{c|ccccc}
Current Trace Type&10 s&6 s&5 s&4 s&2.5 s\\ 
\hline
Noiseless Wireless Rank 1& 94.0 & 94.5 &94.0& 87.5&80.5\\
Noisy Wireless Rank 1& N/A & 87.0&87.5& 87.5&82.0\\
Noiseless Wired Rank 1& 97.0 & 96.0 & 96.5&96.0&88.5\\
Noiseless Wireless Rank 2& 96.0 & 96.5 &97.5& 94.0&88.0\\
Noisy Wireless Rank 2& N/A & 94.0&94.0& 89.5&87.0\\
Noiseless W Wired Rank 2& 99.0 & 97.5 & 98.0 &97.0&93.5\\
 \end{tabular}
 \caption{Rank 1 and rank 2 accuracy (\%) for 1D CNN model when classifying 20 websites with a fully charged iPhone 11.}
  \label{table:r1}
\begin{tabular}{c|ccccc}
Current Trace Types&6 s&5 s&4 s&2.5 s\\ 
\hline
Wireless Rank 1& 95.0 &94.0& 95.5&85.5\\
Wired Rank 1& 74.0& 75.0&70.5&63.0\\
Wireless Rank 2& 97.5 &98.0& 96.5&91.5\\
Wired Rank 2& 83.0& 85.5&82.5&79.0\\
 \end{tabular}
 \caption{Rank 1 and rank 2 accuracy (\%) for 1D CNN model when classifying 20 websites with a fully charged Pixel 4. All traces were collected under normal operation conditions.}
  \label{table:r2}

\end{table}

\subsection{Training and Testing on Different Devices}
In order to see whether or not a cross-device attack is possible in this threat model, we trained the classifier exclusively on current traces from the iPhone and tested on traces from the Pixel and vice versa. When all collected 2.5-second current traces for both devices were input, the classifier was unable to identify traces from the device at all. Training on iPhone traces and testing on Pixel traces resulted in a Rank 1 accuracy of 4.2\% which is worse than a random guess and a Rank 2 accuracy of 12.1\% which is only slightly higher than that of a random guess. Training on Pixel traces and testing on iPhone traces was no better. In this scenario, the classifier achieved a Rank 1 accuracy of 5.7\% and a Rank 2 accuracy of 11.6\%. 

These results align with previous works that found a drop in classification accuracy resulting from training and testing on different devices. This indicates that the information leaked through this side-channel is related to the charging and processor circuitry inside the device and is not directly transferable. An effective realistic attack would likely need to train on traces from a variety of phones in order to be able to generalize and account for more trace variety.

\subsection{Training and Testing on Traces from Different Chargers}
Current traces from a wired, 5 W charger were also collected with both the Pixel and the iPhone. Unlike wireless traces, wired traces from the iPhone were classified with higher accuracy than those of the Pixel. The minimum Rank 1 and Rank 2 accuracies of the classifier on the wired iPhone traces were 88.5\% and 93.5\%, respectively, whereas they were 63.0\% and 79.0\% on the Pixel.

Across all device and charger combinations, our classifier was able to perform well without any preprocessing or changes to architecture. The accuracies achieved by the classifier when trained and tested on wired and wireless traces are similar, indicating that the information leakage from the wireless charging power side-channel is comparable to that of the wired charging power side-channel. In the case of the Pixel 4, the wireless current traces were identified with higher accuracy than the wired current traces.

Figure ~\ref{fig:battery-full} shows the current traces measured using a wired charger while loading zoom.us on iPhone 11. 
Visual comparison suggests that the wired and wireless channels leak the same information when a website is loading; 
the shapes of their power traces when the phone is fully charged are similar. The traces differ, however, because the wireless traces contain a signal with a frequency of approximately 11 Hz and appear to be more noisy in general than the wired traces.

\begin{figure}[!h]
\centering
\subfigure[Training on wireless traces, testing on wired traces.]{\includegraphics{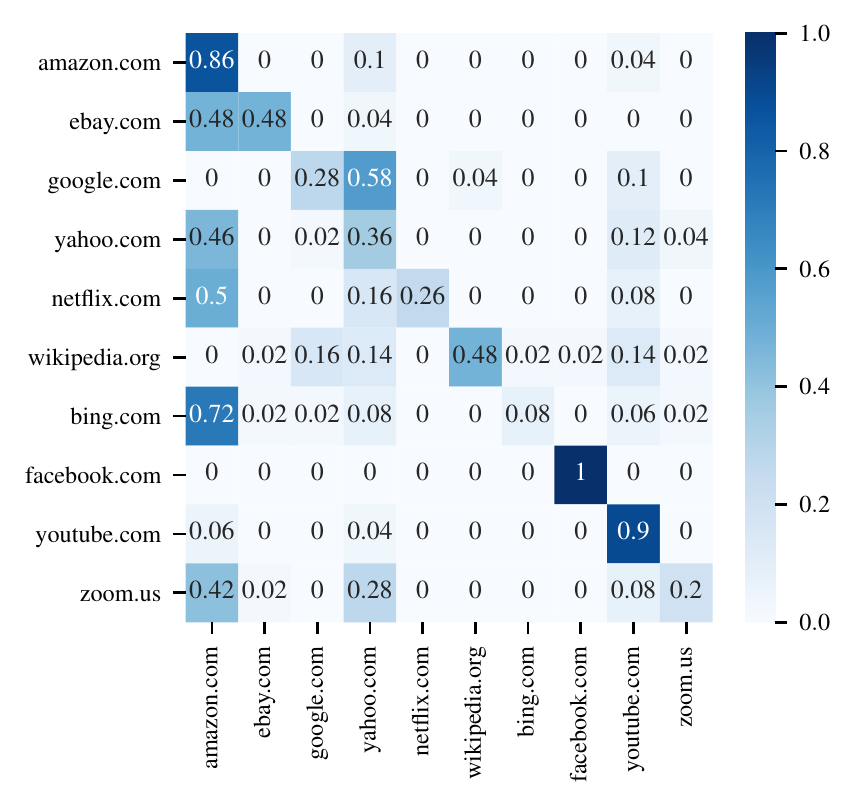}}
\vfill
\subfigure[Training on wired traces, testing on wireless traces.]{\includegraphics{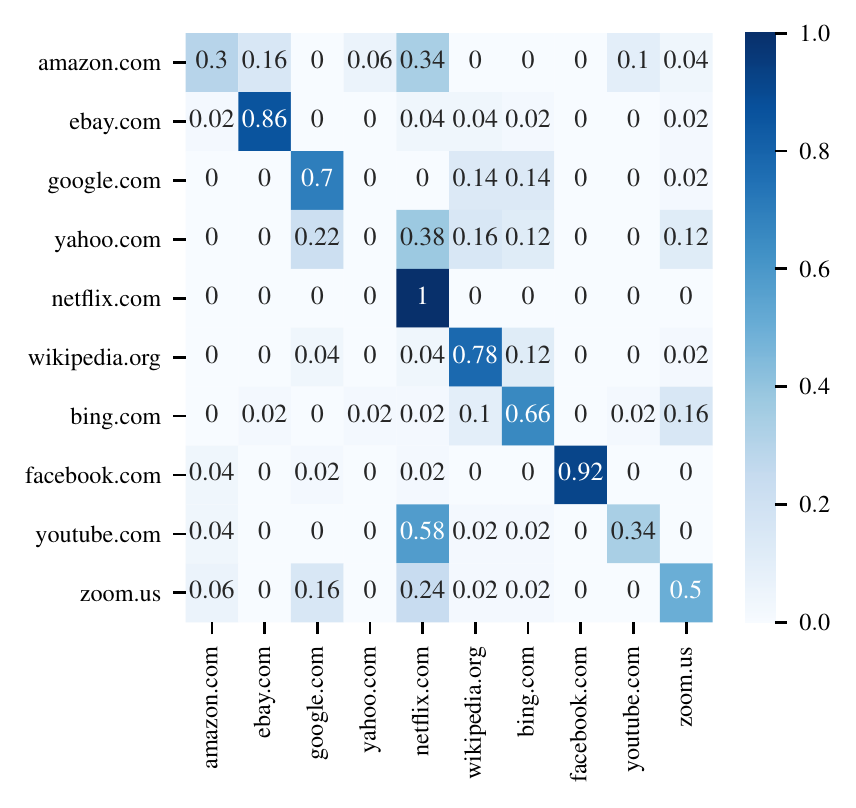}}
\caption{Results from training and testing with different chargers. The vertical axis shows the true label and the horizontal axis shows the predicted label. An ideal classifier would have ones down the diagonal.}
\label{fig:train_wireless}
\end{figure}

In order to measure how comparable both charging side-channels are, the classifier was trained exclusively on current traces from the wireless charger and tested on traces from the wired charger and vice versa. Using 10 websites and 2.5 second long traces, the classifier identified websites correctly with significant accuracy. The results of this experiment are shown in Figure~\ref{fig:train_wireless}. Training on wired traces and testing on wireless traces produced a Rank 1 accuracy of 60.6\% compared to a baseline of 10\% and a Rank 2 accuracy of 75.0\% compared to a baseline of 20\%. Training on wireless traces and testing on wired traces achieved a Rank 1 accuracy of 49.0\%, and a Rank 2 accuracy 68.4\%. The only website that was identified with over 90\% accuracy in both situations was facebook.com.

The existence of cross-channel leakage across both wired and wireless charging indicates that wirelessly charging devices may be susceptible to existing USB power side-channel attacks that have been trained only on wired power data.

\subsection{Impact of Noise}
As evidenced by the results discussed in Section 5.1, the attack is quite resilient to noise, and was able to identify the test traces with high accuracy, even though the circumstances of the device varied between training and testing traces. This demonstrates that our attack is feasible in realistic scenarios where the current trace collected while a website is loading may be corrupted or altered by the existence of other executing processes.

In order to measure how well the attack might perform without noise, current traces were collected from the iPhone 11 while volume, Bluetooth, and cellular data were disabled at a sampling frequency of 500 Hz. Additionally, an outline from the phone was placed over the charger so that the alignment and angle of the phone over the transmitting coil was consistent.

\begin{figure}[!h]
\centering
\includegraphics{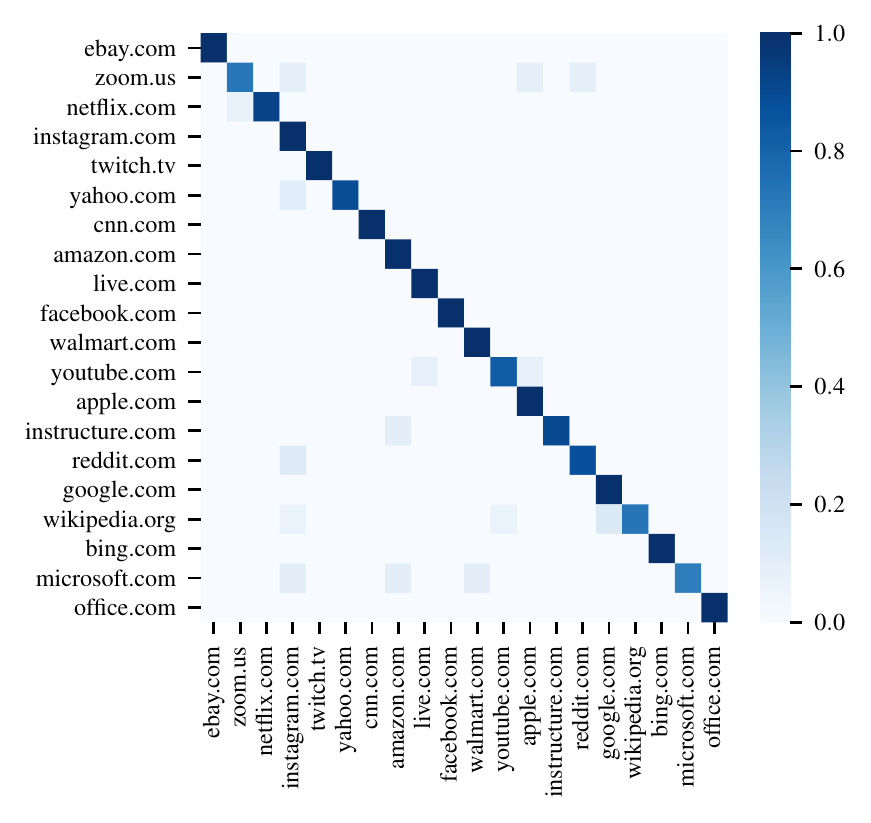}
\caption{Confusion matrix showing the classification of 200 unlabeled current traces.}
  \label{fullcm}
\end{figure}

The classifier performed slightly better when trained and tested on these noiseless traces as opposed to those collected under normal operation conditions. The full results are reported in Table~\ref{table:r1}. When classifying noiseless wireless traces, the classifier obtained a Rank 1 accuracy of at least 80.5\% and a Rank 2 accuracy of at least 88.5\% with trace durations ranging from 2.5 to 10 seconds. We present the confusion matrix for 5-second traces in Figure~\ref{fullcm}.  For comparison, noiseless wired traces collected under the same conditions achieved a Rank 1 accuracy of at least 88.5\% and a Rank 2 accuracy of at least 93.5\%. 

\subsection{Impact of Trace Duration}
In addition to the full 10 or 6 second traces, the classifier was trained and tested on shorter duration traces. These shorter traces were formed by taking a slice of the first $n$ seconds of data from the original trace. Out of all five trace lengths examined, the best wireless and wired rank 1 identification accuracies achieved were with 5-second traces and 6-second traces respectively. While the classifier performed the worst on 2.5-second traces, the overall identification accuracy was still quite high and close to the best rank 1 accuracies out of all trace durations. These shorter traces removed noise present in the full 10-second traces because the websites examined take approximately 4 seconds to load ~\cite{fortune}. However, most websites take over 2.5 seconds to load, so traces of this duration cut off part of the signal from the website loading and therefore deteriorated identification accuracy.
Furthermore, websites that autoplayed videos had consistent leakage in their traces even after they initially loaded.

\subsection{Impact of Length of Time Between Trace Collection and Testing}

In this scenario, training and testing traces were collected on the same iPhone 11 except the test traces were collected nine months after the training traces were. Table~\ref{table:r3} summarizes the results of this scenario. Many of the websites we studied had dynamic content, such as news websites. After many months, the media in these websites completely changed which resulted in the current traces altering as well. Although accuracy was significantly lowered in this experiment, the classifier still performed over four times better than a random guess would achieve.

\begin{table}[!h]
\begin{tabular}{c|ccccc}
Current Trace Type&6 s&5 s&4 s&2.5 s\\ 
\hline
New Traces Rank 1& 18.0 & 20.5& 22.5&13.5\\
New Traces Rank 2& 28.5 & 29.5&33.5&19.5\\ 
 \end{tabular}
 \caption{Rank 1 and rank 2 accuracy (\%) for 1D CNN model when classifying with old training data.}
  \label{table:r3}
\end{table}

\subsection{Impact of Battery State of Charge}

\begin{figure}[!h]
    \subfigure[wirelessly charging (top) and wired charging (bottom)]{\includegraphics[scale=.95]{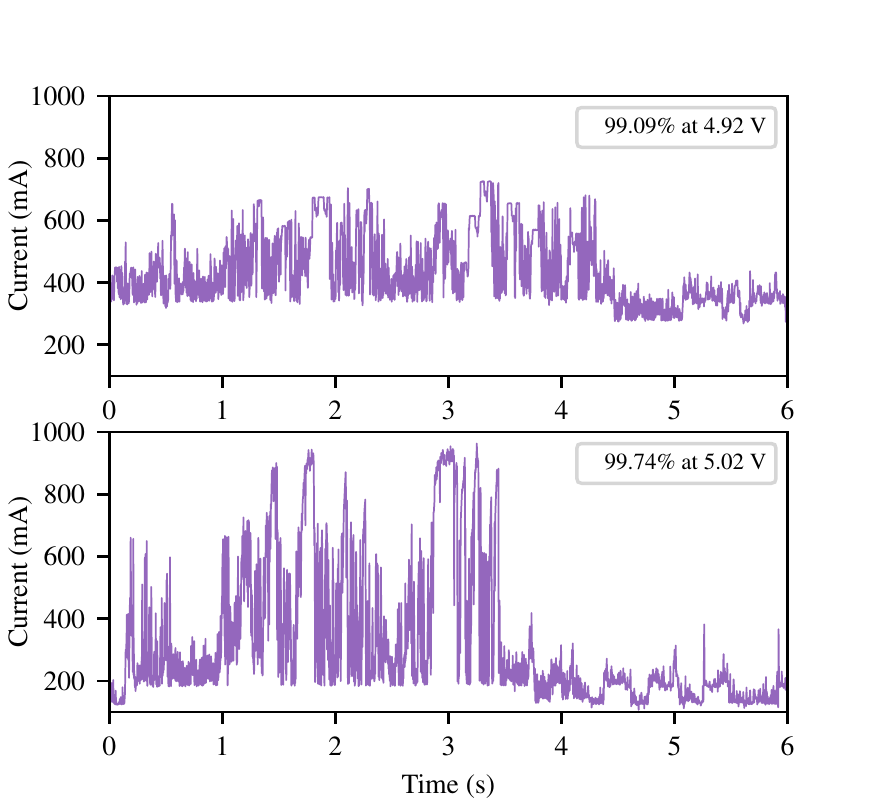}\label{fig:battery-full}}
    \vfill
    \subfigure[76.22\% at 5.10 V]{\includegraphics[scale=.95]{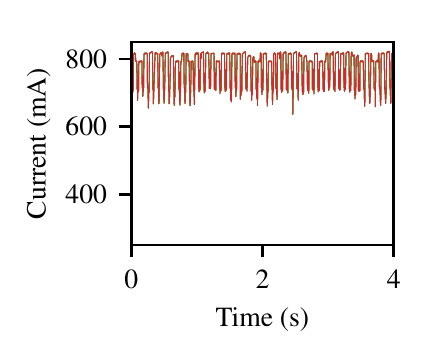}}
    ~
    \subfigure[50.21\% at 4.87 V]{\includegraphics[scale=.95]{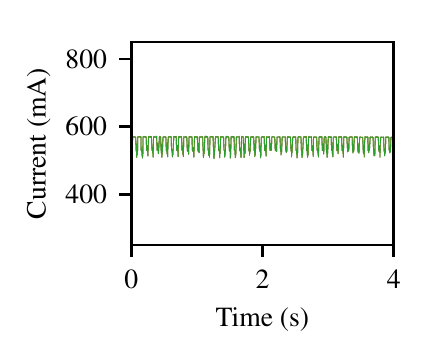}}
    \vfill
    \subfigure[32.59\% at 4.81 V]{\includegraphics[scale=.95]{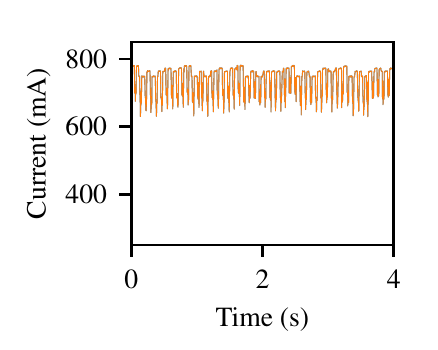}}
    ~
    \subfigure[21.09\% at 4.58 V]{\includegraphics[scale=.95]{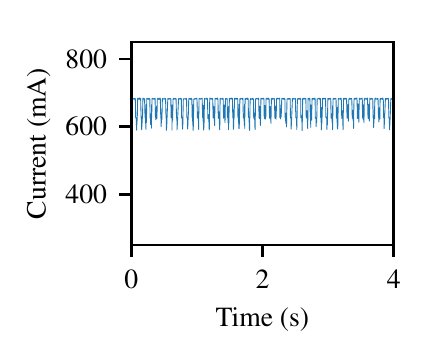}}
    \vfill
    \subfigure[While USB charging, the current traces recorded at the lowest three battery levels are indistinguishable.]{\includegraphics[scale=.95]{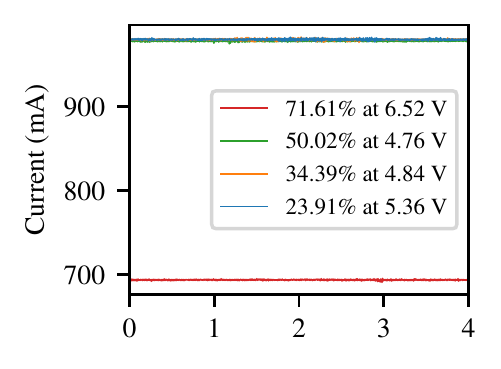}\label{fig:battery-levels}}
    \vfill
    \caption{Comparison of the leakage when loading zoom.us on an iPhone 11. Plots (b)-(e) depict wireless charging.}
    \label{fig:btty}
\end{figure}

Below approximately 80\% state of charge, both wired and wireless charging side-channels observed in this experiment do not leak information and the classifier cannot identify the traces with any significant accuracy. Current traces collected at these states of charge can be seen in Figure~\ref{fig:battery-levels}. For the wired channel, information begins to be revealed when the battery state of charge reaches approximately 95\%. The wireless channel could consistently classify traces with a battery state of charge as low as 90\%.

Figure~\ref{fig:btty} also reveals how the power side-channel through wired charging is affected by the battery level.
The variations from the phone's activities are clearly visible at higher battery levels but not at lower ones.

Previously, Yang et al.\cite{yang-usb} found that power traces collected at battery levels of 30\% were classified with accuracies almost as high as those collected when the battery was fully charged. One explanation for this discrepancy is that the newer smart phones examined in this study have more battery capacity or actively prevent information leakage. This idea is developed further in section 5.5.

In order to further investigate how this side-channel is affected by different battery levels, current traces were collected from two older Apple iPhone models, an iPhone 6s and an iPhone 8, and compared on the same scale. We could collect the power traces for both of these phones using the same data acquisition setup that we used with iPhone 11. 

\begin{figure*}[!t]
\centering
\subfigure[98.35\% at 5.44 V]{\includegraphics[scale=.9]{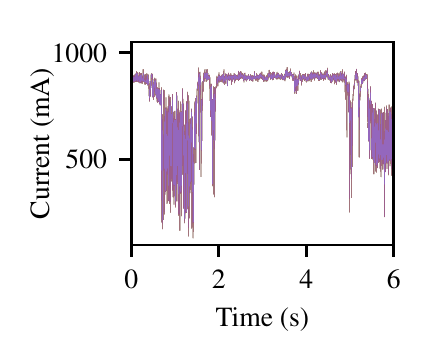}}
\subfigure[70.27\% at 4.68 V]{\includegraphics[scale=.9]{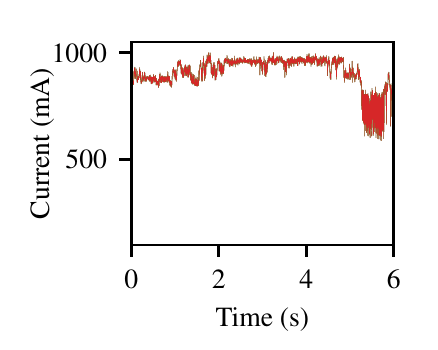}}
\subfigure[51.04\% at 4.70 V]{\includegraphics[scale=.9]{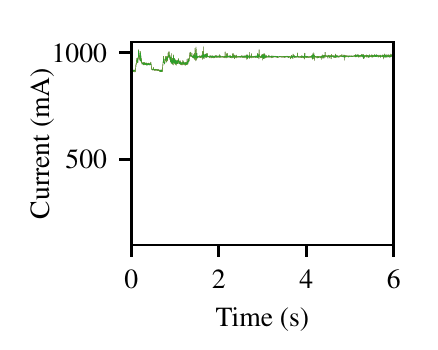}}
\subfigure[31.13\% at 4.69 V]{\includegraphics[scale=.9]{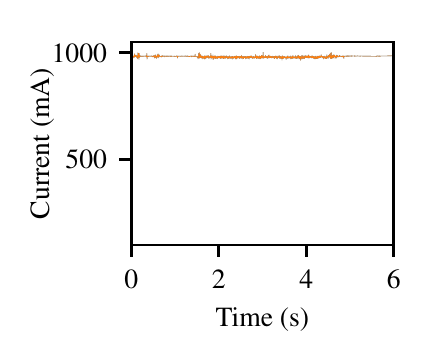}}
\subfigure[21.49\% at 6.93 V]{\includegraphics[scale=.9]{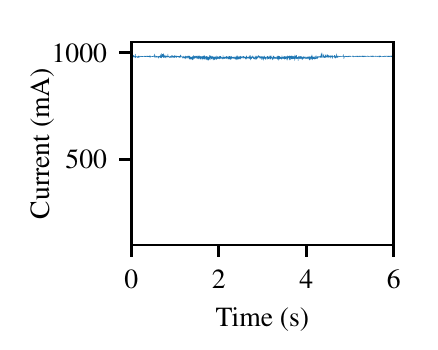}}
\subfigure[98.95\% at 5.05 V]{\includegraphics[scale=.9]{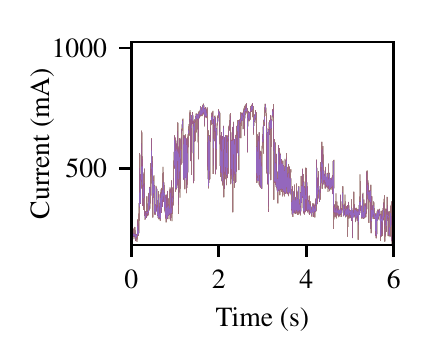}}
\subfigure[70.97\% at 4.72 V]{\includegraphics[scale=.9]{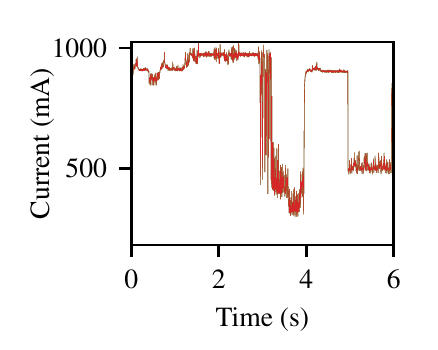}}
\subfigure[52.36\% at 4.70 V]{\includegraphics[scale=.9]{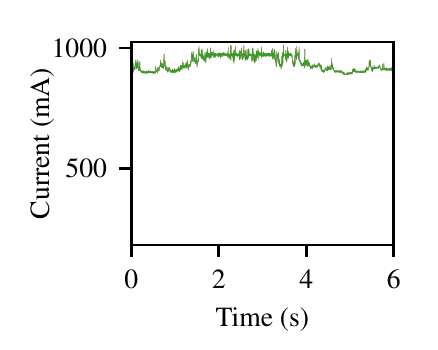}}
\subfigure[31.43\% at 4.70 V]{\includegraphics[scale=.9]{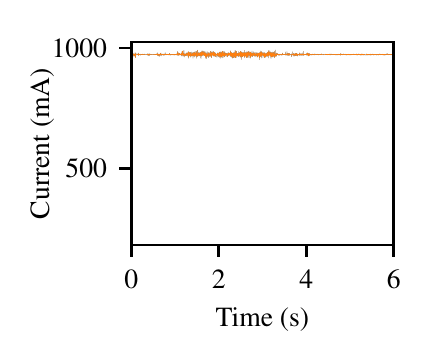}}
\subfigure[19.56\% at 4.66 V]{\includegraphics[scale=.9]{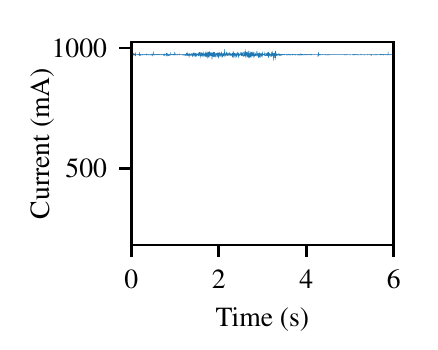}}
\subfigure[98.95\% at 5.01 V]{\includegraphics[scale=.9]{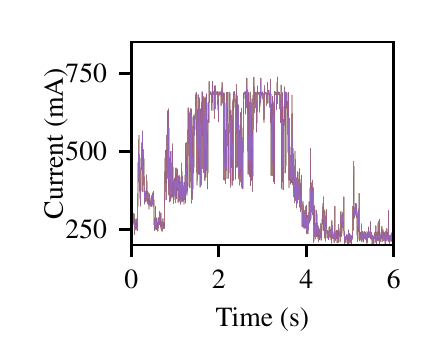}}
\subfigure[70.06\% at 4.91 V]{\includegraphics[scale=.9]{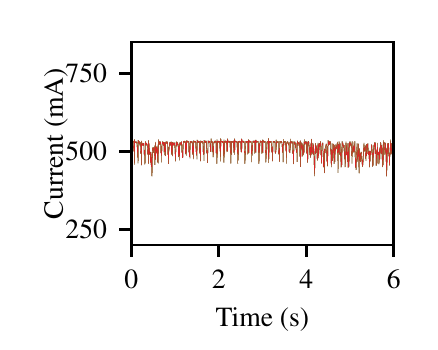}}
\subfigure[53.15\% at 4.88 V]{\includegraphics[scale=.9]{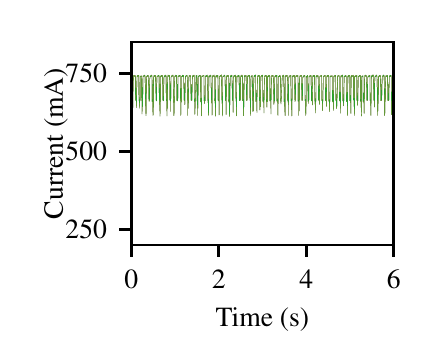}}
\subfigure[31.30\% at 4.92 V]{\includegraphics[scale=.9]{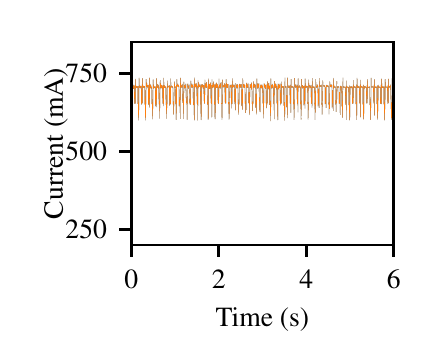}}
\subfigure[19.32\% at 4.66 V]{\includegraphics[scale=.9]{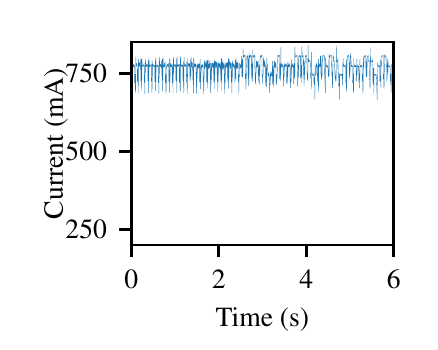}}
\caption{Current traces for loading zoom.us on different devices while wirelessly charging and USB charging: wired iPhone 6s (top), wired iPhone 8 (middle), wireless iPhone 8 (bottom).}
\label{fig:different}
\end{figure*}

Wired traces collected on an iPhone 6s leaked activity at lower states of charge than the iPhone 11 did. This can be seen in Figure ~\ref{fig:different}. Activity was visible at battery levels as low as 50\%, but began to become obfuscated at states of charge 30\% or lower. The iPhone 8 wired power side-channel revealed activity in the same range of battery state of charge as the iPhone 6s.

While the iPhone 6s does not support wireless charging, the iPhone 8 is Qi compatible. Its wireless current traces do not leak any significant information at states of charge less than or equal to 70\% state of charge. It is possible that there was a change in the hardware design between the iPhone 8 and iPhone 11 that removed the USB power side-channel at battery levels below full charge. However, even on the iPhone 8, little activity was revealed at the 30\% battery level compared to the Android phones studied by \cite{yang-usb}. Additionally, even though the iPhone 11 is not as vulnerable to USB power side-channel attacks as the iPhone 8, both phones appear to be similarly susceptible to the wireless charging side-channel attack at higher states of charge.

%% file: countermeasure.tex
\section{Discussions}

\subsection{Countermeasures}

While it enables attacks without a physical connection, the wireless charging side-channel attack is still based on the same secret-dependent variations in the device's power consumption that the traditional power side-channel attacks exploit.
In that sense, the existing countermeasures against power side-channel attacks can also prevent the wireless charging side-channel attack.
For example, Pothukuchi et al.~\cite{maya} show that the power dissipated by a computer can be reshaped to obfuscate the fingerprint left by a running application. Matovu et al.~\cite{matovu} present both a software and hardware solution as defense mechanisms against malicious charging stations. Yan et al. ~\cite{yan-approach} suggest energy obfuscation through code injection, which would embed meaningless code in applications in order to make features in the power trace be less predictable. Similarly, Spreitzer et al.~\cite{spreitzer} proposes execution randomization as a defense mechanism against power analysis attacks. A variety of methods exist to insert random noise into a power trace or obscure sensitive information by making adjustments at the cell level~\cite{popp}. Cronin et al.~\cite{chargesurfing} found that applying a low-pass filter with a cutoff of 60 Hz to collected power trace data reduced the accuracy of their passcode cracking attack to that of a random guess.

To further reduce the amount of information leaked though wireless charging, we may be able to augment the charging algorithm to avoid fully-charging the battery at less trusted locations.
Currently, iPhones running iOS 13 or later employ Optimized Battery Charging, a charging algorithm that reduces the amount of time an iPhone spends fully charged in order to preserve its battery lifespan. This feature uses location data to determine whether or not to delay charging past 80\%~\cite{apple}. If this algorithm could be adjusted to also engage when the iPhone is connected to an untrustworthy charger, then the battery would never leave the constant current Li-ion charging stage as seen in Figure~\ref{fig:long}. Our results show that minimal information would leak to the charger at states of charge less than or equal to this point because the same amount of maximum current will be delivered to the battery regardless of the process currently executing.

\subsection{Other Attacks on Wireless Charging}

This paper investigates the information leakage arising from wireless charging, and
demonstrated the website fingerprinting attack using the wireless power side channel.
The wireless power side channel has a potential to leak many other types of information
or activities on a mobile device that affect the device power consumption.
For example, a recent study~\cite{chargesurfing} showed that the wired USB power
side channel can be used to infer the context of a touchscreen.
It is also an open question if the wireless power side channel can be used to
leak more fine-grained information such as a secret value that is used for processing.
The wireless charging interface may also introduce additional vulnerabilities 
beyond side-channel information leak. For example, a malicious wireless charger may deliver a high current as a way to damage a circuit or perform repeated charging/discharging cycles to reduce battery life.

\subsection{Other Use Cases of Wireless Charging Side-Channel}

Previous studies \cite{watts,lsid} discussed how traditional power side-channels may be used to detect malicious software on embedded devices. 
In a similar fashion, the wireless charging interface may also be leveraged as a way to check the integrity of small mobile or embedded devices without physical connectors, such as a smartwatch.
For such application scenarios, we will need further studies to see if the resolution and the accuracy of the power monitoring through wireless charging is sufficient to detect software changes or malicious activities on an embedded device.

%% file: related.tex
\section{Related Work}

Power analysis attacks are a well established field of research and a variety have been studied in mobile devices. Spreitzer et al.~\cite{spreitzer} presented a thorough categorization system and survey of existing side-channel attacks, especially those applicable to mobile devices. Clark et al.~\cite{clark-identify} found that a computer plugged into a wall was susceptible to an SPA attack and used AC power traces in order to carry out a website fingerprinting attack. While we build upon the existing body of power side-channel and website fingerprinting attacks to demonstrate a vulnerability, our work is the first to identify a wireless charging side-channel which utilizes completely different circuitry than that of wired charging.

Yang et al.~\cite{yang-usb} determined that even when none of a smartphone's data pins are connected, a USB power station can still identify specific activity occurring on the phone. Cronin et al.~\cite{chargesurfing} demonstrate that USB power traces from smartphones leak information about the contents of a device's touch screen. While we also examine this charging power side-channel in our attack, our work differs in several respects. We find that the wireless side-channel is just as, if not more, susceptible to a website fingerprinting attack compared to the traditional wired side-channel. We also sample at 700Hz rather than 250kHz, allowing our attack to be performed by less sophisticated hardware and be more difficult to detect. Additionally, our classifier can effectively classify current traces from different device and charger models without any preprocessing.

The unique combination of hardware and sensor functionality on mobile devices leaves them susceptible to some unique side-channel attacks. Yan et al.~\cite{yan-approach} established a general exploitation approach for a variety of power side-channel attacks on an Android smartphone. While our attack is based on this model, we also demonstrate it on an Apple iPhone and do not require a wired connection, only the physical proximity required to wirelessly charge.

Matyunin et al.~\cite{magnetometer} successfully identify the application running on a phone by studying how CPU operations affect magnetometer measurements. Yang et al.~\cite{yang-fingerprint} showed that the transition between running apps leaves a side-channel in memory that can be used to determine what application was executing. Lifshits et al.~\cite{lifshits} installed a malicious, power monitoring battery in a smartphone in order to identify various types of activity. Qin et al.~\cite{qinetal} also adopt a similar approach to smartphone website fingerprinting by using a malicious application which estimates the fluctuation of power data. The power estimation model employs CPU data that can be accessed without permission in Android 7. In contrast to these works, our work does not require a malicious app or an otherwise compromised phone, because the act of wireless charging itself is vulnerable regardless of permissions set by the operating system.

Another method of website classification besides power side-channels is through traffic and hardware analysis. In contrast to these works, our attack can occur without any software permissions at all. Hintz~\cite{hintz}, Hayes and Danezis~\cite{hayes}, and Lu et al.~\cite{lu} measured the amount of encrypted data being transferred and other metadata to identify webpages even in the face of website fingerprinting defenses. Based on this work, Al-Shehari and Zhioua~\cite{al} proposed a unified traffic analysis attack model for traffic analysis attacks on computers. Our work also examines the Alexa top sites list~\cite{alexa}, but differs in that the side-channel exists locally on the phone's hardware, and is not a result of internet traffic characteristics. Our work demonstrates a novel attack that contributes to the body of website fingerprinting. 

%% file: conclusion.tex
\section{Conclusion}
This paper presents a new side-channel attack that occurs when a Qi-compatible smart device is wirelessly charging and the power consumption of the wireless transmitter is recorded. We show that a low-cost device can be used to collect current traces and infer private information such as browser activity. We demonstrate that this attack can occur even if the user's phone is not fully charged, requires no permission from the phone OS or user, and can occur even if the acquired current trace is quite short (2.5 seconds). Additionally, this new side-channel leaks more information at lower battery levels than a wired power side channel in the same setup.
While this work explores a new side-channel present in all wireless charging compatible smart devices, the entire scope and constraints of the wireless charging side-channel attack and useful countermeasures need to be researched in future work.